\documentclass[journal=acsnano,manuscript=article]{achemso}

\newcommand*{\citen}[1]{%
  \begingroup
    \romannumeral-`\x 
    \setcitestyle{numbers}%
    [\cite{#1}]%
  \endgroup   
}

\usepackage[version=3]{mhchem} 

\author{Joshua Baxter}
\affiliation{University of Ottawa, Department of Physics and Centre for Research in Photonics, Ottawa, Canada}
\email{jbaxt089@uottawa.ca}
\author{Adriana Pérez-Casanova}
\affiliation{Tecnológico de Monterrey, School of Engineering and Sciences, Monterrey, Mexico}
\author{Luis Cortes-Herrera}
\affiliation{Tecnológico de Monterrey, School of Engineering and Sciences, Monterrey, Mexico}
\alsoaffiliation{University of Rochester, Institute of Optics, Rochester, New York, USA}
\author{Antonio Calà Lesina} 
\affiliation{Leibniz Universität Hannover, Hannover Centre for Optical Technologies, Cluster of Excellence PhoenixD, and Fakultät für Maschinenbau (Institut für Transport- und Automatisierungstechnik), Hannover, Germany}
\author{Israel De Leon}
\affiliation{Tecnológico de Monterrey, School of Engineering and Sciences, Monterrey, Mexico}
\alsoaffiliation{University of Ottawa, School of Electrical Engineering and Computer Science, Ottawa, Canada}
\alsoaffiliation{Max Planck–University of Ottawa Centre for Extreme and Quantum Photonics, Ottawa, K1N 6N5 Canada}
\email{ideleon@tec.mx}
\author{Lora Ramunno} 
\affiliation{University of Ottawa, Department of Physics and Centre for Research in Photonics, Ottawa, Canada}
\alsoaffiliation{Max Planck–University of Ottawa Centre for Extreme and Quantum Photonics, Ottawa, K1N 6N5 Canada}
\email{lramunno@uottawa.ca}

\title{Understanding the nonlinear optical response of epsilon near zero materials in the time-domain}


\begin{document}







\begin{abstract}
  The promise of active nanophotonics technology relies on the confinement and control of light at the nanoscale. Confinement via plasmonics, dielectric resonators, and waveguides can be complemented with materials whose optical properties can be
controlled using nonlinear effects. Transparent conducting oxides (TCOs) exhibit strong optical nonlinearities in their near zero permittivity spectral region, on the femtosecond time-scale. Harnessing full control over the nonlinear response requires a deeper understanding of the process. To achieve this, we develop a self-consistent time-domain model for the nonlinear optical response of TCOs and implement it into a three-dimensional finite-difference time-domain code. We compare and tune our simulation tools against
recently published experimental results for intense laser irradiation of thin indium tin oxide (ITO) films. Finally, by simulating intense laser irradiation of ITO-based plasmonic metasurfaces, we demonstrate the full power of our approach. As expected, we find  validating the significant enhancement of the nonlinear response of an ITO-based metasurface over bare ITO thin films. Our work thus enables quantitative nanophotonics design with epsilon-near-zero materials.
\end{abstract}


\section{1. Introduction}
\label{section:intro}

The nonlinear optical response of materials enables a broad range of capabilities within nanophotonics and plasmonic devices, such as extreme refractive index tuning and frequency conversion \citep{boyd_nonlinear_2019}. These capabilities are of utmost importance for applications such as all optical data processing \citep{Cotter1999Nonlinear}, optical switching \citep{Guo2017Solution-Processed,Jiang2018Epsilon-near-zero,Daghooghi2018Ultra-fast}, and active photonics \citep{shaltout_spatiotemporal_2019}. However, the efficiency and scalability of nonlinear optical devices is typically limited by the weak nonlinear response of most natural materials. This results in bulky devices that require long interaction lengths and/or large optical intensities, making it difficult to scale up and integrate nonlinear optical processes in photonic devices with a small footprint. 
\vspace{10pt}

There has been much effort toward the development and understanding of materials with a large nonlinear optical response \citep{Leuthold2010Nonlinear,Kauranen2012Nonlinear,Li2017Nonlinear}. In particular, natural and artificial materials that exhibit a vanishingly small electric permittivity at optical frequencies, known as epsilon-near-zero (ENZ) materials,  promise to be an efficient platform for enhancing  nonlinear optical interactions \citep{Naik2013Alternative,Reshef2019Nonlinear,Kinsey2019Nonlinear,Kinsey2019Near-zero-index}. Enhanced nonlinear processes such as intensity-dependent refractive index \citep{Alam2016Large,Alam2018Large,Lee2018Strong,Caspani2016Enhanced,Carnemolla2018Degenerate,Kinsey2015Epsilon-near-zero} and harmonic generation \citep{Luk2015Enhanced,Capretti2015Enhanced,Capretti2015Comparative,Vincenti2017Second-harmonic}, have been observed in both thin film and nanostructured ENZ materials. A widely studied subset of ENZ materials includes transparent conductive oxides (TCOs), which are degenerately doped semiconductors such as indium tin oxide (ITO) and aluminum-doped zinc oxide (AZO). Their enhanced nonlinear response \citep{Alam2016Large,Lee2018Strong,Caspani2016Enhanced,Luk2015Enhanced,Capretti2015Enhanced}, the modification of their linear optical properties through doping \citep{Noginov2011Transparent}, and their suitability for nanoscale integration through complementary metal-oxide-semiconductor (CMOS) fabrication  \citep{Babicheva2015Transparent}, have positioned these materials as a promising option for enabling nonlinear nanophotonic and nanoplasmonic devices. 
\vspace{10pt}

The nonlinear optical response of TCOs has been extensively studied in their ENZ spectral ranges \citep{Alam2016Large,Caspani2016Enhanced}, where a large nonlinear response has been experimentally observed. This is attributed to intraband transitions occurring in the nonparabolic conduction band of these materials \citep{Liu2014Quantification,Guo2016Ultrafast}. For the case of ITO, different models accounting for the temperature dependence of the plasma frequency \citep{Alam2016Large,Alam2018Large,Guo2016Large,Wang2019Extended}, the effect of free-carrier scattering \citep{Mei2020Role}, other carrier kinetics effects \citep{Secondo2020Absorptive} and nonlocality \citep{Rodriguez-Sune2020Study,scalora_electrodynamics_2020} have been proposed. However, the temporal nature of the nonlinearity is either not accounted for or, when it has been considered, the modelling of the transient response of electron dynamics is separate from the electrodynamics modelling of optical propagation \citep{Alam2016Large,Alam2018Large}. Currently, no model can account for the temporal dynamics of the electron distribution and the optical properties simultaneously and self-consistently as the incident pulse propagates through and interacts with the material. This poses a limitation to describing the time-domain nonlinear response of complex nanophotonic structures that include ENZ materials, thus posing a limitation in their understanding and design.

\vspace{10pt}
In this work, a self-consistent multi-physics numerical model is presented that accounts for the temporal modification of the optical properties and electron distribution in TCO-based ENZ materials, as well fully accounting for electrodynamic propagation. We use a two temperature model (TTM) to describe the electron excitation and thermalization, and account for its effect on the material's optical properties with a temperature-dependent plasma frequency. The implementation of this numerical model in a finite-difference time-domain (FDTD) electrodynamics solver provides access to local optical properties in the time domain. This enables the simulation of the material's nonlinear optical response in space and time, and the design and optimization of complex nonlinear photonic structures incorporating TCO-based ENZ materials
\vspace{10pt}

 
The structure of the article is as follows. We introduce our material model for the optical nonlinearity in Section 2\ref{section:s2} and provide an intuitive explanation of the physical processes behind the temporal change in the optical properties. In Section 3\ref{section:s3}, we refine and validate our model by reproducing experimental results from experiments published in Ref. \citen{Alam2016Large}, where we explore the effect of the pump pulse’s central wavelength, incident angle, and intensity on the nonlinear optical response of ITO thin films. Comparisons between the simulation results and experimental data demonstrate quantitative agreement. In Section 4\ref{section:s4}, we investigate the time-domain nature of the nonlinearity, and compare our findings to pump-probe experiments published in Ref. \citen{Alam2016Large}. There, we also present movies of our simulations that show the evolution of the electric field amplitudes for both pump and probe simulations. In Section 5\ref{section:s6}, we simulate an ITO-based plasmonic metasurface, demonstrating the versatility of the FDTD implementation of our model. We find that the presence of the plasmonic particle enhances the refractive index change by a factor of five over a bare ITO thin film. Finally, in Section 6\ref{section:s7}, we conclude our work.

\section{2. Model for TCO nonlinear optical response}
\label{section:s2}

Transparent conductive oxides (TCOs) are wide-bandgap, degenerate semiconductors that possess a metallic optical response in the near infrared (NIR) due to an abundance of free electrons in the conduction band. Therefore, the frequency-domain permittivity of TCOs is well described in the NIR spectral range by the Drude model \citep{Wang2019Extended}

\begin{equation}
\label{Drude}
    \varepsilon\left(\omega\right)=\varepsilon_\infty-\frac{\omega_p^2}{\omega^2+i\gamma\omega},
\end{equation}

\noindent where $\omega_p=\sqrt{Ne^2/\varepsilon_0m^\ast}$ is the plasma frequency, $\gamma$ is the damping coefficient, $\varepsilon_\infty$ is the high-frequency permittivity, $\omega$ is the angular frequency of the electromagnetic field,  $m^*$ is the effective mass of the conduction electrons, $e$ is the charge of an electron, $N$ is the free electron density, and $\varepsilon_0$ is the permittivity of free space.  Note that according to Eq. \ref{Drude}, the wavelength at which the real part of the permittivity equals zero (the zero-crossing wavelength) is given by $\lambda_{\rm ZC}= 2 \pi \sqrt{\varepsilon_{\infty}}c/ \omega_p$.
\vspace{10pt}

The properties of the conduction electrons are dictated by the conduction-electron energy band. Typically, the parabolic band approximation can be applied, such that $m^*$ does not change upon interaction with light. This is generally not the case in TCOs exposed to intense, ultrafast laser pulses. Due to their high doping concentration (and thus high carrier density), a correction term is required in the dispersion relation \citep{Pisarkiewicz1990Nonparabolicity},

\begin{equation}
\label{nparabolic}
    \frac{\hbar^2k^2}{2m_0^\ast}=E+CE^2,
\end{equation}

\noindent where $E$ is the electron energy, $\hbar k$ is the magnitude of the crystal momentum, $\hbar$ is the reduced Planck’s constant, $m_0^*$ is the effective mass at the conduction band minimum, and $C$ is a non-parabolicity factor. This non-parabolic band structure results in a plasma frequency that depends on conduction band electron temperature, $T_e$. An expresion for $\omega_p (T_e)$  is derived in the Supplementary Information (SI) Section 1, where we use the semiclassical Boltzmann equation in the relaxation-time approximation, assuming an isotropic conduction band. The resulting formula is 

\begin{equation}
\label{PF}
    \omega_p^2\left(T_e\right)=\frac{e^2}{3m_0^\ast\varepsilon_0\pi^2}\left(\frac{2m_0^\ast}{\hbar^2}\right)^{3/2}\int_{0}^{\infty}{dE\ {(E+CE^2)}^{3/2}\left(1+2CE\right)^{-1}\left(-\frac{\partial f_{FD}\left(E,T_e\right)}{\partial E}\right)},
\end{equation}

\noindent where $f_{FD}=(\exp{((E-\mu)/k_B T_e)} + 1)^{-1}$ is the Fermi-Dirac distribution, $\mu$ is the chemical potential (discussed later) , and $k_B$ is the Boltzmann constant. This temperature dependence of the plasma frequency is a consequence of the energy dependence of the electronic effective mass in the conduction band \citep{Wang2019Extended}. Note that a different derivation of Eq. \ref{PF} is given in Ref. \citen{Guo2016Ultrafast}, resulting in the same equation.


\vspace{10pt}

\begin{figure}
\centering
\includegraphics[width = 3.5in]{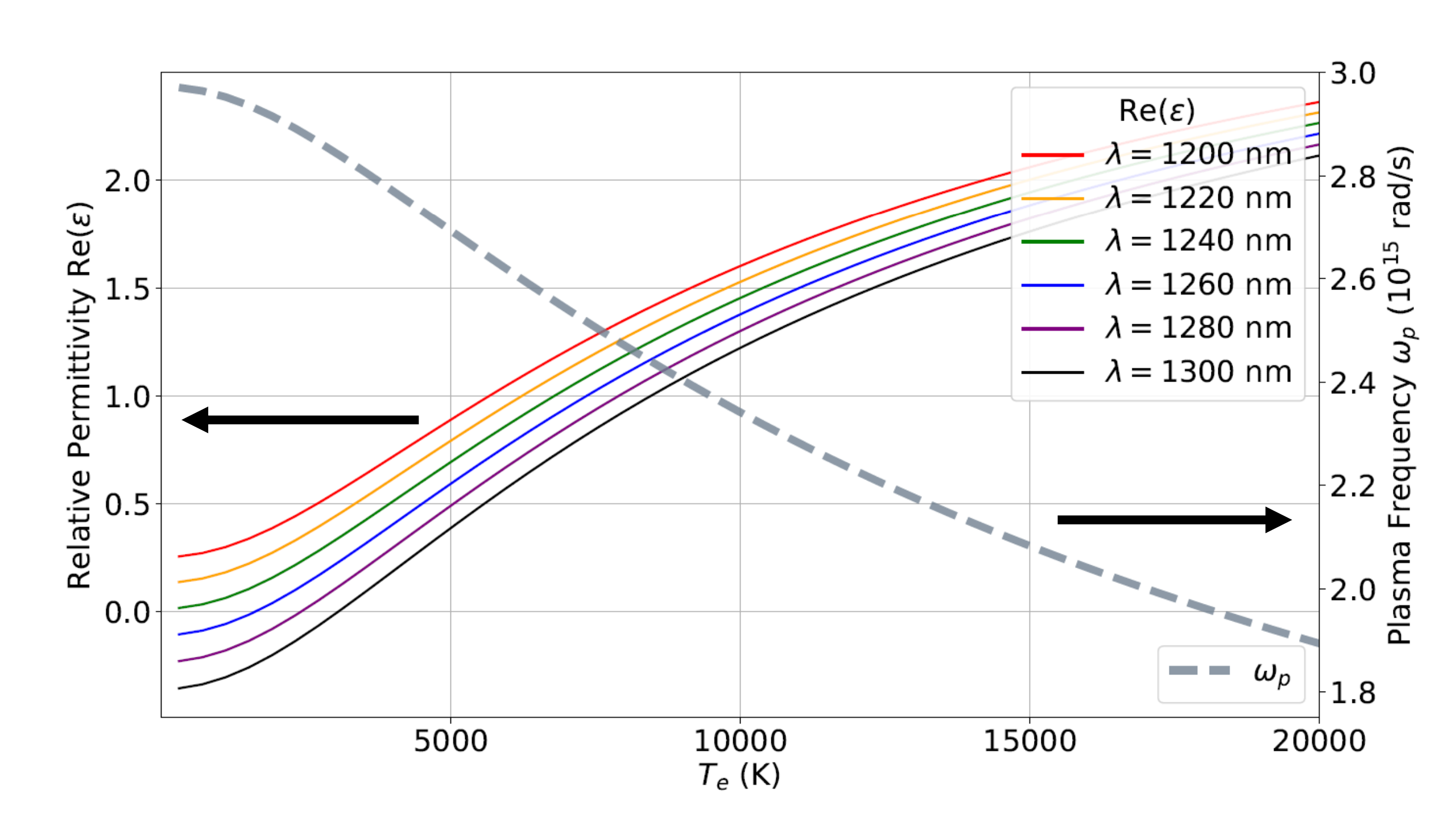}
\caption{Left vertical axis: Real component of the relative permittivity of ITO plotted as a function of electron temperature for wavelengths ranging from 1200 nm (top solid curve) to 1300 nm (bottom solid curve), calculated using Eq. \ref{Drude}. The parameters used for this plot are given in Section 3. Right vertical axis: Plasma frequency plotted as a function of temperature (dashed line), calculated via numerical integration of Eq. \ref{PF}.}
\label{pffig}
\end{figure}

In Fig. \ref{pffig}, we plot both the plasma frequency (evaluated via numerical integration of Eq. \ref{PF}) and the resulting real component of the Drude permittivity (Eq. \ref{Drude}) for several wavelengths as a function of electron temperature, $T_e$. The plasma frequency decreases monotonically with $T_e$, leading to a dramatic increase in the relative permittivity. For wavelengths between 1200 nm and 1300 nm, room-temperature (300 K) ITO is an ENZ material. In fact, for $\lambda > 1240$ nm, room-temperature ITO behaves as a metal since Re($\varepsilon$) $<0$, but as the ITO heats, Re($\varepsilon$) becomes positive, and it transitions to a dielectric. We discuss the dynamics of this metal-to-dielectric transition further in Section 4\ref{section:s4}.

\vspace{10pt}

\begin{figure}
\centering
\includegraphics[width = \linewidth]{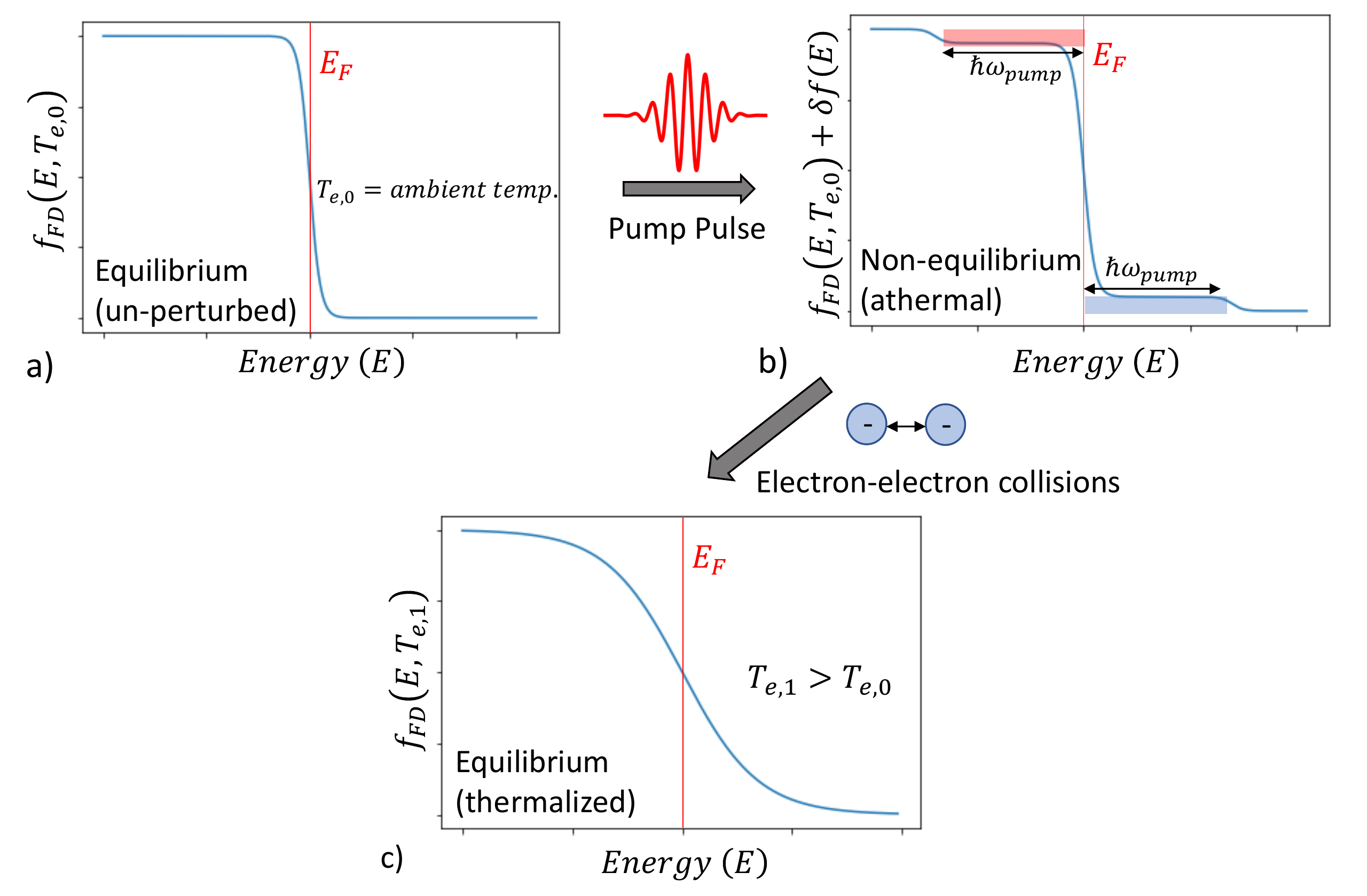}
\caption{Qualitative illustration of the electron distribution thermalization process. a) Initially the material is in thermal equilibrium with the ambient medium and is described by a Fermi-Dirac distribution $f_{FD}(E,T_{e,0})$ at ambient temperature $T_{e,0}$. b) The pump pulse acts by exciting electrons below the Fermi level  in the range  ($E_F-\hbar\omega_{pump}$) to $E_F$ (red shading) to states above the Fermi level, $E_F$ to ($E_F+\hbar\omega_{pump}$) (blue shading). This can be described as a perturbation added to the Fermi-Dirac distribution $\delta f(E)$. c) Through electron-electron collisions, the distribution thermalizes to a hot Fermi Dirac distribution with temperature $T_{e,1} > T_{e,0}$.}
\label{physics}
\end{figure}

Before introducing our model for the electron thermalization dynamics, we first qualitatively describe the effect on the material's electron distribution upon irradiation by an intense pulse with a center angular frequency of $\omega_{pump}$. As with other degenerate semiconductors, the Fermi energy $E_F$ is dependent on the doping concentration, though it typically lies in the conduction band. Defining the zero-energy level to be at the conduction band edge, the Fermi level of ITO is on the order of 1 eV inside the conduction band. As illustrated in Fig. \ref{physics}, when a pump pulse is incident on TCOs, conduction band electrons with energies $E$ between $E_F-\hbar \omega_{pump}$ and $E_F$ are excited above the Fermi level leading to an athermal (nonequilibrium) distribution of electrons (Fig. \ref{physics}b) that is perturbed from the original Fermi-Dirac distribution (Fig. \ref{physics}a) \citep{Voisin2001Ultrafast}. Subsequent electron-electron scattering results in the athermal energy being rapidly redistributed in the electron gas, resulting in a heated Fermi distribution with temperature $T_e$ (Fig. \ref{physics}c).

\vspace{10pt}

Now we turn to our deriving our quantitative model. The thermalized electron and lattice subsystems are treated using an extended two temperature model (TTM) \cite{Carpene2006Ultrafast} given by

\begin{subequations}
\label{TTM_Full}
\begin{align}
\frac{\partial U(t,\mathbf{r})}{\partial t}&=-\frac{U(t,\mathbf{r})}{\tau_{ee}(T_e)}-\frac{U(t,\mathbf{r})}{\tau_{ep}(T_e,T_l)}+P\left(\mathbf{r},t\right),\\
    C_e\left(T_e\right)\frac{\partial T_e(t,\mathbf{r})}{\partial t}&=-g_{ep}\left(T_e,T_l\right)\left(T_e\left(t,\mathbf{r}\right)-T_l\left(t,\mathbf{r}\right)\right)\ +\frac{U\left(t,\mathbf{r}\right)}{\tau_{ee}\left(T_e\right)}, \\
    C_l\left(T_l\right)\frac{\partial T_l\left(t,\mathbf{r}\right)}{\partial t}&=g_{ep}\left(T_e,T_l\right)\left(T_e\left(t,\mathbf{r}\right)-T_l\left(t,\mathbf{r}\right)\right)+\frac{U\left(t,\mathbf{r}\right)}{\tau_{ep}\left(T_e,T_l\right)},
\end{align}
\end{subequations}

\noindent  where $T_l$ is the lattice temperature, $C_e$ is the electron heat capacity, $C_l$ is the lattice heat capacity, $g_{ep}$ is the electron-phonon coupling coefficient, $\tau_{ee}$ is the electron-electron scattering time, $\tau_{ep}$ is the electron-phonon scattering time, and $P$ is the electromagnetic power absorbed by the conduction electrons. In Eqs. \ref{TTM_Full}, the electron and lattice subsystems are treated separately, and energy is exchanged between the electron and lattice subsystems via electron-phonon coupling $g_{ep}$. This version is referred to as the extended TTM since it accounts for the athermal electron energy density $U(\textbf{r},t)$  (J/m\textsuperscript{3}), a scalar field which acts as a thermal energy source to the conduction electrons. 

\vspace{10pt}

Due to the relatively low Fermi energy of ITO, the electron-electron scattering time $\tau_{ee}$ is on the order of a few femtoseconds (becoming sub-femtosecond for a hot electron gas) as we describe in the SI Section 2.  Because time scales of interest in optical simulations are typically on the order of hundreds of femtoseconds (much larger than $\tau_{ee}$), the effect of including $U$ in our modelling is negligible. For simplicity we thus use a simplified version of the TTM,

\begin{subequations}
\label{TTM}
\begin{align}
C_e\left(T_e\right)\frac{\partial T_e(t,\mathbf{r})}{\partial t}&=-g_{ep}\left(T_e,T_l\right)\left(T_e\left(t,\mathbf{r}\right)-T_l\left(t,\mathbf{r}\right)\right)\ +P\left(\mathbf{r},t\right),\\
C_l\left(T_l\right)\frac{\partial T_l\left(t,\mathbf{r}\right)}{\partial t}&=g_{ep}\left(T_e,T_l\right)\left(T_e\left(t,\mathbf{r}\right)-T_l\left(t,\mathbf{r}\right)\right).
\end{align}
\end{subequations}

\noindent The parameters $C_e$, $C_l$, and $g_{ep}$ are all temperature dependent; expressions for calculating them are given in SI Section 3, along with a method for calculating the temperature-dependent chemical potential $\mu$.
\vspace{10pt}

In order to complete our model, Maxwell’s equations are solved in the time domain concurrently with the Drude model via

\begin{gather}
    \mu_0\frac{\partial\mathbf{H}}{\partial t}=-\mathrm{\nabla}\times\mathbf{E} \label{HM},\\
{\varepsilon_\infty\varepsilon}_0\frac{\partial\mathbf{E}}{\partial t}=\nabla\times\mathbf{H}-\mathbf{J} \label{EM},\\
\frac{\partial\mathbf{J}}{\partial t}=-\gamma\mathbf{J}+\varepsilon_0\omega_p^2(T_e)\mathbf{E}, \label{Current}
\end{gather}

\noindent where $\textbf{E}$ is the electric field, $\textbf{H}$ is the magnetic field, and $\textbf{J}$ is the conduction electron current density that accounts for intraband transitions. For constant $\omega_p$, Eq. \ref{Current} is equivalent to the Drude model (Eq. \ref{Drude}). As there are no interband transitions in the NIR for ITO, we use a constant permittivity to describe the effects of the bound electrons that is absorbed into $\varepsilon_{\infty}$. 

\vspace{10pt}

The TTM and electrodynamics equations are coupled via two terms: (i) the absorbed power $P=\textbf{J}\cdot\textbf{E}$, given by Poynting's theorem, and used as the source in Eq. \ref{TTM}a; and (ii) the plasma frequency $\omega_p(T_e)$, calculated via Eq. \ref{PF} and Eqs. \ref{TTM}, and used as the source of the nonlinearity in the electrodynamics equations. Solved together, Eqs. \ref{PF}, \ref{TTM} and \ref{HM} - \ref{Current} describe the time-dependent, thermally driven optical nonlinearity in TCOs.

\vspace{10pt}

As mentioned in the introduction, a more accurate treatment would involve including hydrodynamic terms in Eq. \ref{Current} and a nonlinear model for the bound electrons as in Ref. \citen{Baxter_parallel_2021}. After some numerical experiments using our existing hydrodynamics solver\citep{bin-alam_hyperpolarizability_2021}, we found that including hydrodynamic terms has very little effect on the nonlinearity when solved in conjunction with the TTM.
\vspace{10pt}

We use an in-house parallel 3D-FDTD code \cite{Lesina2015On,Baxter_parallel_2021} to solve Eqs. \ref{HM} - \ref{Current}. In the same code, we also solve the TTM (Eqs. \ref{TTM}) using the Runge-Kutta method \citep{DeVries2011first}. At each step in the FDTD simulation, the electric field, magnetic field, current density, and the electron and lattice temperatures are updated self-consistently.

\section{3. Model Validation and Refinement in ITO thin films}
\label{section:s3}


To validate the model presented in the previous section, we compare it against experimental results obtained by femtosecond laser irradiation of ITO thin films. Although there have been many experimental studies in recent years \citep{Alam2016Large,Alam2018Large,Liu2014Quantification,Wang2019Extended,Secondo2020Absorptive,Rodriguez-Sune2020Study,Zhou2020Broadband}, due to the sensitivity of ITO’s optical properties to manufacturing specifications, we focus on reproducing the results of the article by Alam, De Leon and Boyd~\citep{Alam2016Large}, in which the effect of the pump pulse on the strength of the nonlinearity in 310 nm ITO thin films is investigated using several tests. First, the authors measure the effect of the pump central wavelength and incident angle on the nonlinear refractive index of ITO. Second, they show how the reflection, transmission, and absorption shift as a function of the pump intensity. Third, they measure the time-domain dynamics of the nonlinearity via pump-probe measurements of the transmission. We use our model to reproduce the first two of these tests in this section, and tackle the third (time-domain dynamics) in the next section. Through these comparisons, we are able to further refine our model to most accurately match this particular set of  experimental data. We also perform an optimization procedure to determine optimal pulse parameters for maximizing the nonlinearity. 
\vspace{10pt}

We consider a 310 nm layer of ITO on glass, as in Ref. \citen{Alam2016Large}. To model the one-dimensional thin-film geometry  (along \textit{y}), we apply Bloch periodic boundary conditions to the \textit{x} and \textit{z} directions of our three-dimensional (3D) simulation domain enabling us to use oblique-incident pulses \citep{Yang2007simple,Aminian2006Spectral}, which we set to be \textit{p}-polarized. We use the room-temperature Drude model parameters for ITO reported in Ref. \citen{Alam2016Large}, $\omega_p=2\pi\times 473$ THz, $\gamma=0.0468~\omega_p$, and $\varepsilon_\infty=3.8055$, which result in a zero-crossing wavelength of $\lambda_{\rm ZC} = 1237$~nm.

\vspace{10pt}

Ref. \citen{Alam2016Large} quantifies the nonlinear response by assuming it is Kerr-like with an intensity dependent refractive index $n=n_0+n_2 I$. The real part of the effective nonlinear index, Re$(n_2)$, and the effective nonlinear attenuation constant, $\beta=(4\pi/\lambda) \hspace{3pt}$ Im$(n_2)$, are extracted via \textit{z}-scan measurements. It was observed that the maximum Re$(n_2)$ and minimum $\beta$ are found when the pump wavelength is centered at 1240 nm and the angle of incidence is 60-degrees with respect to the ITO surface normal.

\vspace{10pt}

 Treating the thermal nonlinearity described in Section 2\ref{section:s2} as Kerr-like is a rough approximation for two primary reasons: (i) there is no reason to assume that higher odd-order terms do not contribute to the nonlinearity, and (ii) the nonlinearity is not instantaneous. Although our numerical implementation allows us to calculate the nonlinear polarization (or current density) field using more formal definitions, to compare with the experimental results, we calculate an effective $n_2$ from the simulation data. Though there are several ways to go about this, we use the following computationally efficient method. We calculate the mean conduction electron temperature ($\overline{T_e}$) within the ITO over space and time. This is used to obtain an effective plasma frequency and thus, by applying Eq. \ref{Drude}, an effective, temperature dependent, refractive index $n(\overline{T_e})$ = Re($\sqrt{\varepsilon}$), which differs from the room temperature refractive index $n_0$  by $\overline{\Delta n}=n(\overline{T_e})-n_0$. We then extract an effective $n_2$ using

\begin{equation}
    n_2 = \frac{\overline{\Delta n}}{\overline{I}},
\end{equation}

\noindent where $\overline{I}$ is the time-averaged pulse intensity. The true refractive index $n(\textbf{r},t)$ will vary in space and time during pulse irradiation and therefore $\Delta n(\textbf{r},t)=n(\textbf{r},t)-n_0$ also varies in space and time. While the time and space dependence of $n_2$ is not measured in Ref. \citen{Alam2016Large}, we do explicitly calculate $\Delta n(\textbf{r},t)$ in Section 5\ref{section:s6} and SI Section 4.

\vspace{10pt}

\begin{figure}
\centering
\includegraphics[width = \linewidth]{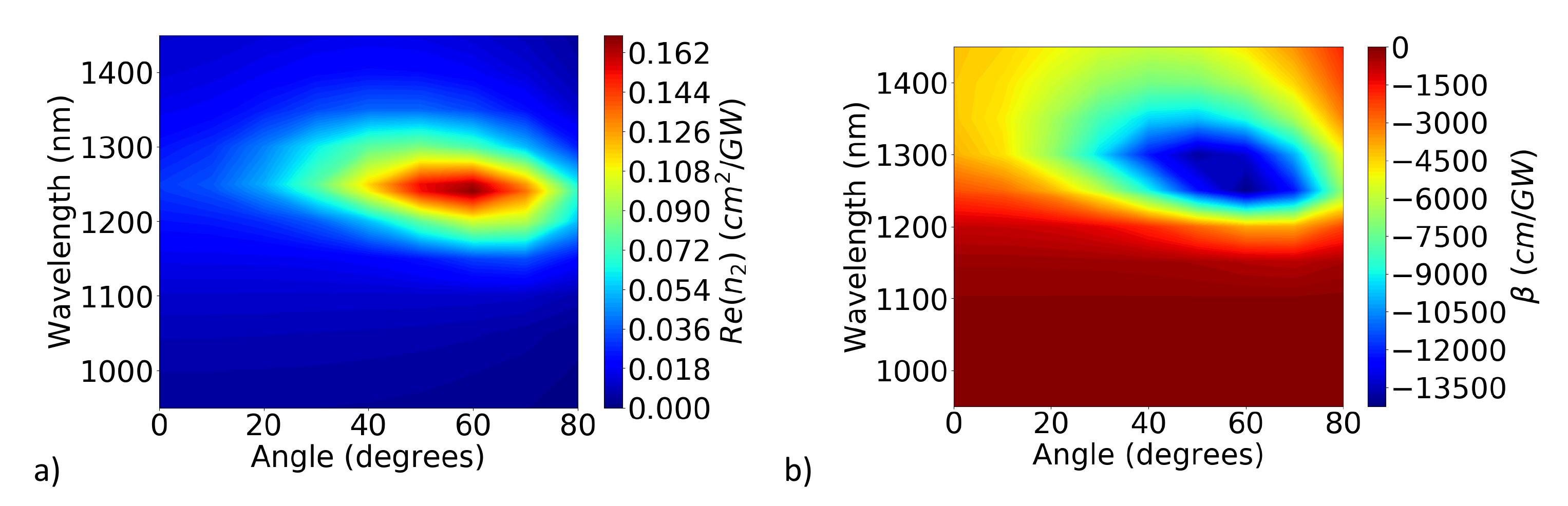}
\caption{Angle (horizontal axis) and central wavelength (vertical axis) dependence of the a) real part of the nonlinear refractive index and b) the attenuation constant of an ultrafast pulse incident on a 310 nm ITO thin film on glass.}
\label{fig1}
\end{figure}

We carry out a two-dimensional parameter sweep of incident angle (in 10-degree increments) and wavelength (in 50 nm increments) and plot the effective $n_2$ and $\beta$ values in Fig. \ref{fig1}. Our results indicate that $n_2$ is maximized at a wavelength of 1250 nm and an angle of 60 degrees, in agreement with the experimental results. Further, our values of $n_2\approx 0.16$~cm\textsuperscript{2}/GW and $\beta\approx-13000$ cm/GW are in reasonable quantitative agreement with the values obtained in Ref. \citen{Alam2016Large}, which are $n_2\approx 0.11$~cm\textsuperscript{2}/GW and $\beta\approx-7500$~cm\textsuperscript{2}/GW.

\vspace{10pt}

\begin{figure}
\centering
\includegraphics[width = 3in]{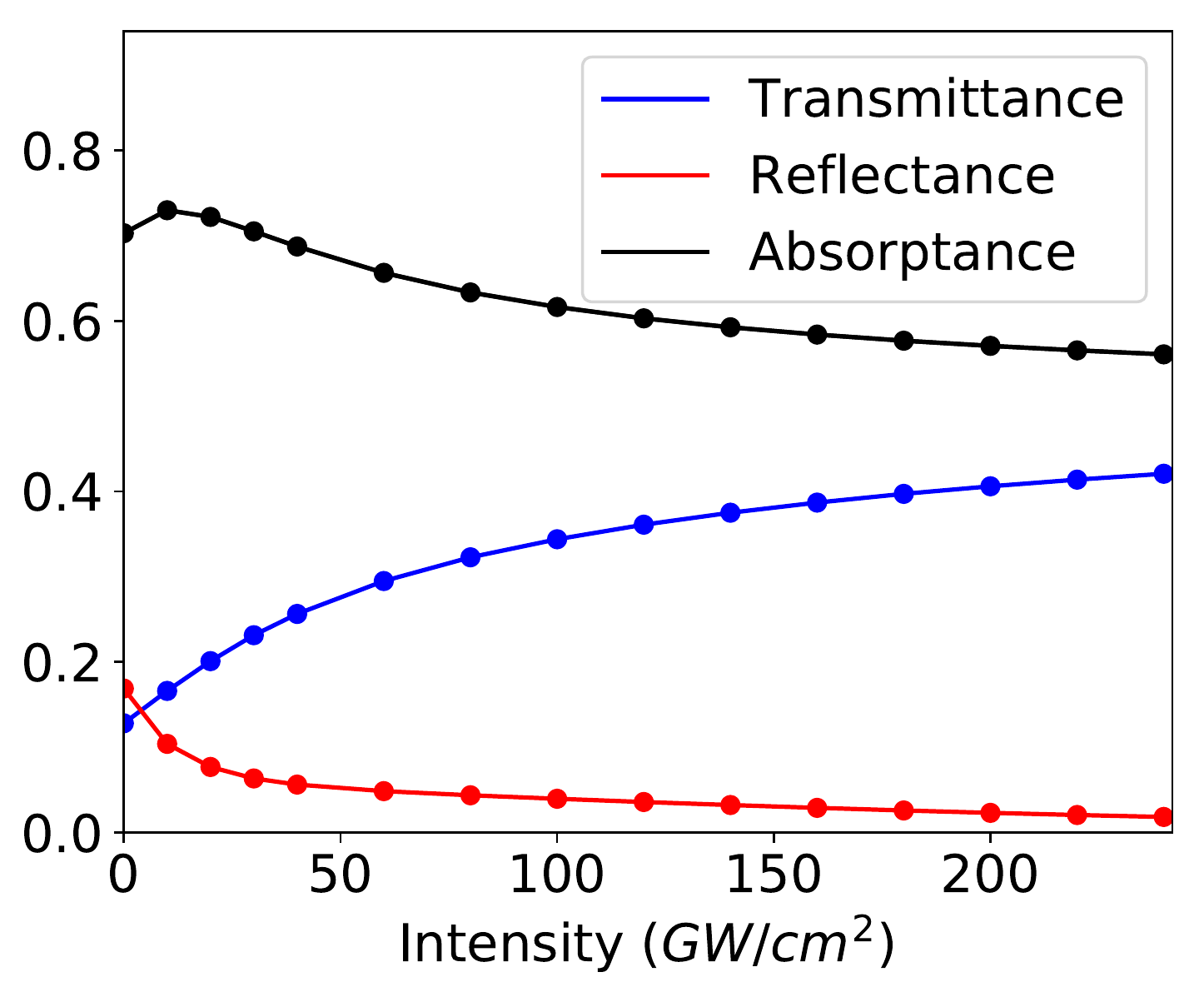}
\caption{Intensity dependence of the transmittance (blue), reflectance (red), and absorptance (black) of a 1240 nm pump incident at a 30 degree angle from vacuum onto a 310 nm ITO thin film on glass.}
\label{fig2}
\end{figure}

Next, we determine the effect of the incident pulse intensity on the reflectance, transmittance, and absorptance for a central pump wavelength of 1240 nm and an incident angle of 30 degrees as in Ref. \citen{Alam2016Large}. We show the results of our calculations in Fig. \ref{fig2} where we see an increase in the transmittance from $T\approx 0.13$ to $T\approx 0.4$ as the intensity increases from 0 to 250 GW/cm\textsuperscript{2}, and a corresponding decrease in the reflectance from $R\approx 0.19$ to $R\approx 0.02$. Our results agree quantitatively with the experimental measurements \cite{Alam2016Large} which show an increase in the transmittance from $T\approx 0.12$ to $T\approx 0.4$ and a corresponding decrease in reflectance from $R\approx 0.2$ to $R\approx 0.04$ for the same intensity range.

\vspace{10pt}

A feature of note in Fig. \ref{fig2} is the saturation of the nonlinearity for high intensities. For simulations in which we assume that the Drude damping coefficient $\gamma$ is temperature independent, this saturation occurs for intensities larger than 250 GW/cm\textsuperscript{2}. In order to produce Fig. \ref{fig2}, and reach agreement with experiments, we needed to consider a model wherein the damping coefficient increases with temperature. As a first-order approximation, we use

\begin{equation}
    \gamma(T_e)=\gamma_0 \Big(1+\frac{T_e}{T_0}\Big),
\end{equation}

\noindent where $\gamma_0$ is the low-intensity Drude damping coefficient and $T_0=15000$ K is a fitting parameter with units of temperature. The increase in the Drude damping coefficient is reasonable because we would indeed expect a decrease in electron mobility due to an increase in electron-phonon, electron-impurity, and grain boundary collisions.  This effect was also found experimentally in Ref. \citen{Wang2019Extended}, where an almost-linear relationship was found between an effective $\gamma$ and the pulse intensity by fitting the measured spectra to the Drude model. The fact that $\gamma$ depends on the temperature highlights the benefit of time-domain modelling as we can extract simple, yet effective models for optical parameters that otherwise would require semi-classical or quantum mechanical calculations.

\vspace{10pt}

Finally, we optimize properties of the pump pulse (center wavelength, incident angle, and linear chirp for a given intensity) in order to maximize the change in the refractive index $\Delta n(\textbf{r},t)$.  Using a Nelder Mead algorithm, we find that for lower intensities, $\Delta n$ is largest when the center wavelength is tuned to the ENZ wavelength. Interestingly, for higher intensities, a larger center wavelength is required to maximize $\Delta n$. For all intensities, the optimal incident angle is $\theta_{inc}\approx 30^{\circ}$ and the linear chirp has negligible effect. These results are presented and further explained in SI Section 4.

\section{4. Analysis of time-domain dynamics}
\label{section:s4}
Part of the interest in ITO’s strong nonlinearity stems from its ultrafast response time, limited mainly by the pulse width (since the athermal relaxation time is small), meaning that we can have precise control of the instantaneous refractive index. Degenerate pump-probe measurements have shown that the rise time is less than 200 fs for a 150 fs laser pulse and the fall time was approximately 360 fs \citep{Alam2016Large}. In this section, we present corresponding pump-probe simulations using a two-step numerical process. A primary FDTD-TTM simulation computes the temperature fields as a function of space and time as the material is excited by a pump pulse. These temperature fields are stored and subsequently used in separate probe simulations, where it is assumed that the probe intensity is low enough that it has no effect on the temperature field.

\vspace{10pt}

\begin{figure}
\centering
\includegraphics[width = 2.5in]{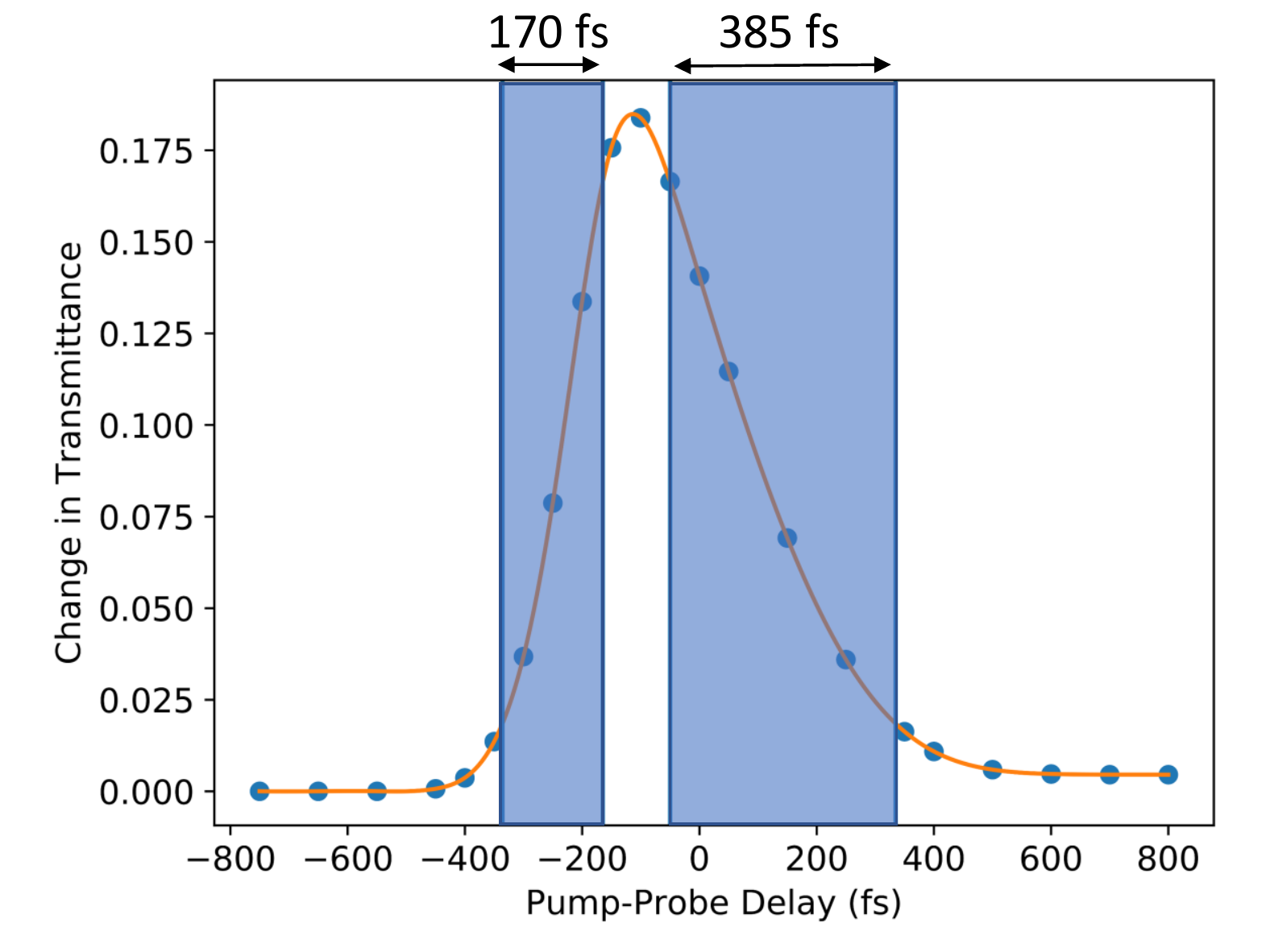}
\caption{Transmittance of a probe pulse through a 310 nm film of ITO on glass previously irradiated with an ultrafast pump pulse as a function of pump-probe delay. The pump is a 1240 nm pulse at an incident angle of 30 degrees.}
\label{fig3}
\end{figure}

We plot the transmission at $\lambda=1240$ nm as a function of probe pulse delay in Fig. \ref{fig3}, which shows a rise time of 170 fs and a fall time of 385 fs. As in Ref. \citen{Alam2016Large} these are defined as the time it takes for the transmittance to rise (fall) from 10\% (90\%) to 90\% (10\%) of its maximum change. In order to obtain Fig. \ref{fig3}, we needed to adapt a model for the electron-phonon coupling strength $g_{ep}$ that was originally derived for metals (Eq. \ref{EPC}) and contains a free parameter. Since the fall time of the optical response is controlled by $g_{ep}$, we compared numerical calculations to the experimental results to obtain the free parameter; this is presented in SI Section 3.

 \vspace{10pt}

\begin{figure}
\centering
\includegraphics[width = \linewidth]{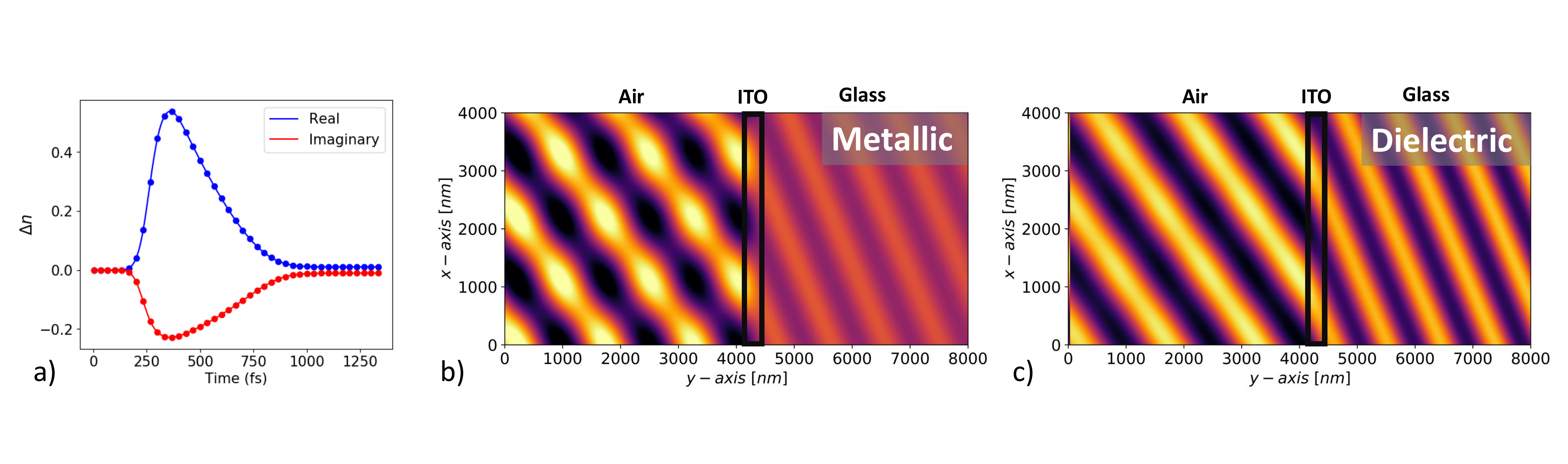}
\caption{a) Change in the effective refractive index of a 310 nm ITO film in the same simulation as Fig. \ref{fig3}. b) Snapshot of electric field amplitude early in the pulse when the ITO film is metallic, and c) later in the pulse when the ITO has become dielectric. The incident pulse propagates from the vacuum (left) through the ITO film into the glass substrate (right).}
\label{fig7}
\end{figure}

 In Fig. \ref{fig7}a we show the corresponding effective refractive index of the ITO film as a function of time, which is extracted using an effective medium theory where we consider the thin film to be an effective (uniform) medium \citep{smith_determination_2002}. We see that the real part of the refractive index increases by approximately 0.5 and the imaginary part decreases by approximately 0.2. Such an index change suggests a transition from metallic (Re$[\varepsilon]<0$) to dielectric (Re$[\varepsilon]>0$) behavior in the epsilon-near-zero region, which also explains the large increase in transmittance in Fig. \ref{fig3}.

 \vspace{10pt}

We illustrate this further in Fig. \ref{fig7}b and \ref{fig7}c, which shows snapshots of the electric field amplitude during the simulation of a 150 fs pulse incident on an ITO film on glass. We see in Fig. \ref{fig7}b that, early in the simulation, a large amount of the incident light is reflected, as evidenced by the interference patterns, indicating that ITO behaves like a metal. Fig. \ref{fig7}c shows the electric field amplitude later in the same pulse where the ITO layer is hotter and there is much less reflection (no discernible interference pattern), indicating that ITO is behaving as a dielectric.

 \vspace{10pt}

All of our results above show that the coupled FDTD-TTM model for ITO agrees remarkably with experiments. The results of this section further indicates the strength of the FDTD implementation. Because our modelling is in the time domain, we can track the nonlinear dynamics as a function of space and time. In the Supplementary Materials (SM), we present  10 sets of movies showing the evolution of the pump electric field amplitude for different incident angles, center wavelengths, and intensities. Also included are the corresponding videos of the electronic temperature field evolution. These are briefly discussed in SI Section 5. 

\vspace{10pt}

Here we highlight one example set of movies. \textit{Movie1} shows the electric field amplitude evolution of a 250 GW/cm\textsuperscript{2} intensity pump, centered at 1240 nm, incident at 30 degrees, and  \textit{Movie1-T} shows the corresponding electronic temperature field evolution. In \textit{Movie1}, we see that at low temperatures (before the pump peak), the ITO film exhibits metallic properties as it reflects much of the incident pulse and absorbs much of the transmitted light, resulting in an evanescent wave in the ITO layer. Indeed, the angle of transmission in the ITO is small due to the low refractive index. As the pulse grows in intensity (and $T_e$ increases), we see the skin depth of the ITO increase (Im($\Delta n$)$<$0), the angle of transmission in the ITO increase (Re($\Delta n$)$>$0), and a corresponding increase in transmission into the glass substrate. At this point ITO is behaving like a dielectric. Finally, near the end of the pulse, we see the interference pattern in vacuum diminish significantly due to the decreased reflectance.

\section{5. Nonlinear response of a plasmonic metasurface on ITO}
\label{section:s6}

In the previous sections, we refined and validated our model using experimental measurements published in Ref. \citen{Alam2016Large} using both spectral and time domain results. It was found that the pump's wavelength and angle of incidence have a strong effect on the magnitude of the refractive index change. Intuitively, one might suppose that the strength of the  nonlinear optical response can be further increased by enhancing the near fields within ITO films by including nanostructures. Plasmonic nanoparticles have been used to enhance optical processes of all sorts, including, nonlinear optical effects \citep{Gaponenko2020Colloidal}. Since our model is built into FDTD, we are not restricted to simulations of simple planar geometries, but we can simulate the refractive index shift in  ITO in the presence of near-by plasmonic particles, where the 3D spatial structure and dispersion of the plasmonic material are taken into account self consistently.

\vspace{10pt}

As an illustrative example, we consider a plasmonic metasurface consisting of a periodic array of gold rectangular nanoparticles on a 310 nm thick layer of ITO on glass. The nanoparticle has a length of 500 nm, width of 300 nm, and thickness of 40 nm, and is centered in a 600 nm $\times$ 600 nm unit cell. This geometry was chosen because its has a very large absorption peak ($A\approx$ 0.9) within the frequency range of interest; this peak occurs at $\lambda_{peak} = $ 1350 nm. We use Bloch boundary conditions in our FDTD simulations, where a plane wave that is linearly polarized along the length of the particle, is incident on an infinitely periodic metasurface at an angle of 30 degrees. As in SI Section 4, we use a 2 GW/cm\textsuperscript{2}, pump pulse which allows for ease of comparison to our results there.

 \vspace{10pt}

\begin{figure}
\centering
\includegraphics[width = 4.7in]{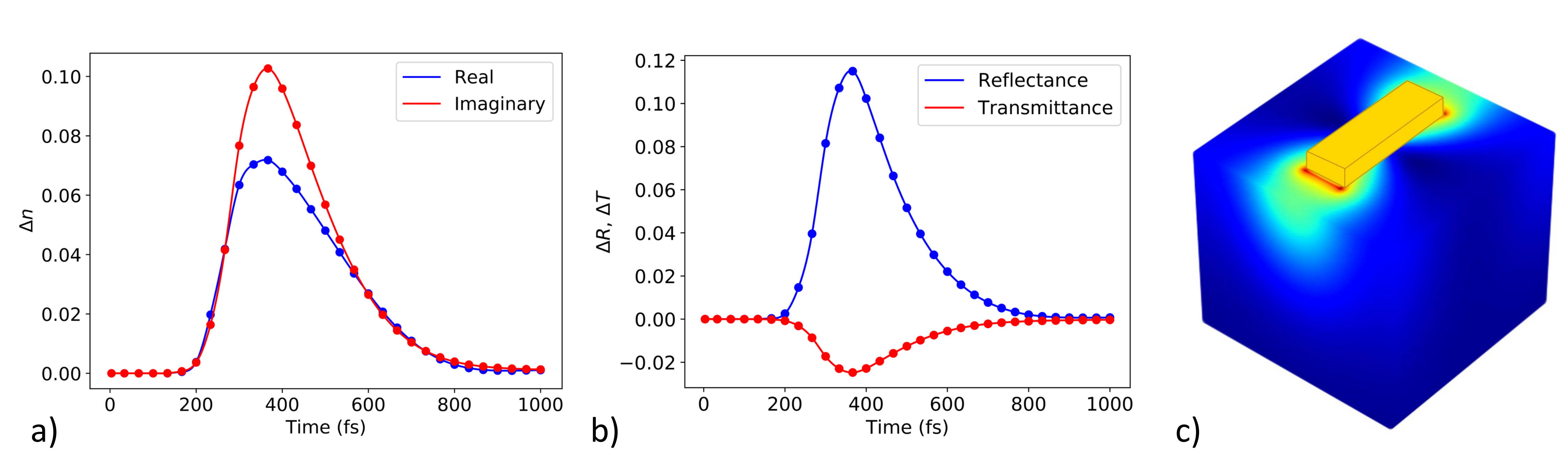}
\caption{a) Enhanced change in refractive index and b) change in reflectance and transmittance of an ITO-based plasmonic metasurface. c) 3-D time-domain snapshot of the temperature field in the ITO film at the peak of the laser pulse, with the plasmonic particle shown on top.}
\label{fig8}
\end{figure}

We calculate the instantaneous effective index change of the metasurface (using effective medium theory \cite{smith_determination_2002} as described above) as a function of time, which is plotted in Fig. \ref{fig8}a. We find that the maximum of $\Delta n$ is now five times larger than for the bare ITO thin film (SI Section 4, Fig. \ref{optimization}b) irradiated at the same incident pump intensity. In Fig. \ref{fig8}b we plot the change in reflectance and transmittance, $\Delta R$ and $\Delta T$, respectively. We find that the transmittance decreases during the pulse irradiation ($\Delta T < 0$), in contrast to the increase that was demonstrated for bare ITO thin films ($\Delta T > 0$) in the previous section (Fig. \ref{fig3}) and SI Section 4 (Fig. \ref{optimization}c). In addition to this change in sign, the magnitude of $\Delta T$ is also two times larger than that seen in Fig. \ref{optimization}c in SI Section 4. The large enhancement in $\Delta T$ is due to the plasmonic particle, which creates electromagnetic hotspots when excited near the localized surface plasmon resonance. As their names imply, these hotspots are regions of enhanced Joule heating leading to a stronger nonlinearity being induced in the ITO film. By comparing Fig. \ref{fig8}b to Fig. \ref{optimization}c we also observe a longer transient nonlinear response when the plasmonic particle is present. This occurs because the decay time of the plasmon resonance is longer than the decay time of the ITO nonlinear response.

\vspace{10pt}
Here, it is worth noting that the effective medium theory does not just output an effective refractive index $n=\sqrt{\varepsilon \mu}$, but also an effective admittance $Y=\sqrt{\varepsilon/\mu}$. Typically, for  sufficiently small feature sizes, we expect these values to be equal as long as the material is non-magnetic ($\mu=1$). We have verified that this indeed remains true for bare ITO films considered in the previous sections. However, this is not the case for the metasurface investigated in this section, where applying effective medium theory suggests an effective magnetic response of the metasurface. In fact, we can see this effect in Fig. \ref{fig8} where we see an increased imaginary component of the effective index (Fig. \ref{fig8}a), however, there is a decrease in the absorptance (Fig. \ref{fig8}b), which can only occur if $\mu\neq 1$. This effective magnetic behaviour is due to complex response of the plasmonic particle, and will change the ENZ behaviour, because we must take $\mu$ into account. This will be investigated in future work.
 
\vspace{10pt}

Previous experimental work has investigated ultrafast irradiation of ITO-based plasmonic metasurfaces \citep{Alam2018Large}. There, it was also found that the transmittance decreases during pulse irradiation for the metasurface but increases (with a smaller line-width) for a bare ITO film. This further validates the results of this (and the previous) section. The magnitudes of $\Delta T$  and $\Delta n$ are not directly comparable between our work and that of Ref. \citen{Alam2018Large}, as a different ITO sample was used there, with different optical properties and a different pump intensity. However, the enhancement of $\Delta n$ when a plasmonic particle is present versus a bare ITO film is comparable.

 \vspace{10pt}
 
 We also present movies showing the time evolution of the electric field amplitude and temperature field of the plasmonic metasurface on ITO while being irradiated by an oblique incidence pump pulse. In \textit{Movie11}, we see the incident pulse exciting the plasmonic particle and creating electromagnetic hotspots at the particle edges. These result in electronic temperature hotspots as shown in \textit{Movie11-T} and Fig. \ref{fig8}c. As the pump pulse peaks, we start to see strong interference between the incident and reflected pulse in the vacuum region due to the increased reflectance (as also indictated by the $\Delta R$ plot in  Fig. \ref{fig8}b). There is very little decrease in the transmittance to balance the reflectance ($\Delta T$ plot in Fig. \ref{fig8}b) which, in \textit{Movie11}, is evident by the increased penetration of the field into the ITO film, suggesting an increased skin depth during intense field irradiation.

\section{Conclusion}
\label{section:s7}

 We introduced a self-consistent time-domain numerical model for the intensity-dependent refractive index of TCO-based ENZ materials. The nonlinear optical response of TCOs is described by a Drude-like permittivity where the plasma frequency is dependent upon electron temperature. A two-temperature model describing the electron temperature distribution in time and space as induced by irradiation with an ultrafast intense laser pulse was implemented directly into a three dimensional finite-difference time-domain electrodynamics solver. Our self-consistent numerical method captures the full nonlinear spatio-temporal dynamics in TCO-based ENZ materials, including those within complex geometries containing, for example, plasmonic nanostructures in close proximity. We validated our implementation via comparisons with experimental results for a bare ITO film, where we obtained quantitative agreement, as well as an ITO film decorated with plasmonic nanoantennas. We also demonstrated the use of the model for optimizing the strength of the optical nonlinearity over pump-pulse parameters. This approach to modeling the nonlinear optical response of TCO-based ENZ materials will be instrumental for understanding the complex dynamics of hot-electron nonlinearity of TCOs and for designing sophisticated nonlinear nanophotonic devices with femtosecond-scale response times.

\newpage
\section*{Supplemental Information}
\label{section:si}
\subsection*{Section 1: Temperature dependence of plasma frequency}

In this section, we derive a model for the temperature dependence of the plasma frequency. We begin with the Boltzmann transport equation, which describes the time evolution of the conduction band non-equilibrium distribution function $f(\textbf{r},\textbf{k},t)$, where $\textbf{r}$ and $\textbf{k}$ are the three-dimensional position vector and wave-vector respectively, and where $f(\textbf{r},\textbf{k},t)  d\textbf{r}d\textbf{k}/4\pi^3$  is the number of conduction electrons in phase space from $\textbf{r}$ to $\textbf{r} + d\textbf{r}$ and $\textbf{k}$ to $\textbf{k} + d\textbf{k}$. As we will see, our final result is equivalent to that given by Ref. \citen{Guo2016Ultrafast}.

In equilibrium, $f(\textbf{r},\textbf{k},t)$ is given by the Fermi-Dirac distribution $f_{FD}$ and thus deviations from equilibrium are quantified by the athermal distribution

\begin{equation}
\label{delta}
    \delta f(\textbf{r},\textbf{k},t)=f(\textbf{r},\textbf{k},t)-f_{FD},
\end{equation}

\noindent whose dynamics in an external force $\textbf{F}$ are described by the semi-classical Boltzmann transport equation \cite{Ashcroft1976Solid}

\begin{equation}
\label{BE}
    \frac{\partial\delta f}{\partial t}+\textbf{v}_\textbf{k}\cdot\frac{\partial\delta f}{\partial \textbf{x}}+\textbf{F}\cdot\frac{1}{\hbar}\frac{\partial f}{\partial \textbf{k}}=\Big(\frac{\partial\delta f}{\partial t}\Big)_{collisions},
\end{equation}

\noindent where $\textbf{v}_\textbf{k}=\hbar^{-1}\partial E/\partial \textbf{k}$  is the group velocity of a Bloch electron in the conduction band (where $E$ is its energy), and where the last term describes the effects of electronic collisions. For the latter, we use the relaxation-time approximation

\begin{equation}
    \Big(\frac{\partial\delta f}{\partial t}\Big)_{collisions}=-\frac{\delta f}{\tau},
\end{equation}

\noindent where $\tau$ is the inverse scattering rate. To simplify the equation further, we will assume that $\partial f/\partial\textbf{k} = \partial f_{FD}/\partial\textbf{k}$ in the third term on the left hand side of Eq. \ref{BE}.

We take the external force to be $\textbf{F}=-e\textbf{E}$ , where $\textbf{E}$ is the local electric field. Since the optical wavelength is much larger than the lattice spacing, we disregard the spatial variation of $\delta f$. Further, using, $\frac{\partial f_{FD}}{\partial \textbf{k}}=\hbar\textbf{v}_\textbf{k}\frac{\partial f_{FD}}{\partial E}$, and setting $\gamma=1/\tau$, Eq. \ref{BE} becomes

\begin{equation}
    \frac{\partial\delta f}{\partial t}-e\textbf{E}\cdot\textbf{v}_\textbf{k}\frac{\partial f_{FD}}{\partial E}=-\gamma \delta f,
\end{equation}

\noindent which in the frequency domain gives 

\begin{equation}
   (-i\omega+\gamma)\delta f =e\textbf{E}\cdot\textbf{v}_\textbf{k}\frac{\partial f_{FD}}{\partial E}. 
\end{equation}

We  calculate the current density $\textbf{J}$ by integrating $e\textbf{v}_\textbf{K}$ over all $\textbf{k}$ states weighted by $\delta f(\textbf{k})/(4\pi^3)$ \cite{Ashcroft1976Solid}

\begin{equation}
    \textbf{J}=-\frac{e}{4\pi^3}\int d^3k~\delta f(\textbf{k})\textbf{v}_\textbf{k}=-\frac{e^2}{4\pi^3}\int d^3k~ \textbf{v}_\textbf{k}\Big(\textbf{E}\cdot\textbf{v}_\textbf{k}\frac{\partial f_{FD}}{\partial E}\Big) (-i\omega+\gamma)^{-1},
\end{equation}

\noindent from which we identify the conductivity tensor

\begin{equation}
\label{sigma1}
    \sigma = -\frac{1}{-i\omega+\gamma}\frac{e^2}{4\pi^3}\int d^3k~\textbf{v}_\textbf{k}\textbf{v}_\textbf{k}\frac{\partial f_{FD}}{\partial E},
\end{equation}

\noindent where $\textbf{v}_\textbf{k}\textbf{v}_\textbf{k}$ is a dyadic tensor defined by $(\textbf{v}_\textbf{k}\textbf{v}_\textbf{k})_{ij}=(\textbf{v}_\textbf{k})_i(\textbf{v}_\textbf{k})_j$, where $i$ and $j$ refer to Cartesian coordinates.

Under the assumption of an isotropic band structure, $\textbf{v}_\textbf{k}=\hbar^{-1}\nabla_\textbf{k} E$ is parallel to $\textbf{k}$ and therefore, in spherical coordinates $\textbf{v}_\textbf{k}=(\sin\theta \cos\phi,\sin\theta \sin\phi,\cos\theta) v_k$, where $v_k$ is the magnitude of the velocity vector, thus giving

\begin{equation}
    \textbf{v}_\textbf{k}\textbf{v}_\textbf{k}=\begin{pmatrix}
\sin^2\theta\cos^2\phi & \sin^2\theta\cos\phi\sin\phi & \sin\theta\cos\theta\sin\phi\\
\sin^2\theta\cos\phi\sin\phi & \sin^2\theta\sin^2\phi & \sin\theta\cos\theta\sin\phi \\
\sin\theta\cos\theta\cos\phi & \sin\theta\cos\theta\sin\phi & \cos^2\theta
\end{pmatrix} v_k^2=\textbf{M}(\theta,\phi)v_k^2.
\end{equation}

Rewriting the integral in Eq. \ref{sigma1} in spherical coordinates, we obtain

\begin{equation}
    \sigma=-\frac{1}{-i\omega+\gamma}\frac{e^2}{4\pi^3}\int_0^{2\pi}\int_0^\pi \textbf{M}(\theta,\phi)\sin\theta d\theta d\phi\int dk~k^2v_k^2\frac{\partial f_{FD}}{\partial E}.
\end{equation}

\noindent By integrating over $\theta$ and $\phi$, we find that the non-diagonal terms vanish, and the diagonal terms are all equal to $\frac{4}{3} \pi$, thus confirming that the conductivity is isotropic, given by

\begin{equation}
    \sigma=-\frac{1}{-i\omega+\gamma}\frac{e^2}{3\pi^2}\int dk~k^2v_k^2\frac{\partial f_{FD}}{\partial E}.
\end{equation}

\noindent Now we use Eq. \ref{nparabolic} from which we can determine expressions for $k$ and $v_k$. Changing the integration variable from momentum magnitude $k$ to energy $E$, using Eq. \ref{nparabolic} in the main text, we obtain

\begin{equation}
\label{sigma}
    \sigma=-\frac{1}{-i\omega+\gamma}\frac{e^2}{3m_0^*\pi^2}\Big(\frac{2m_0^*}{\hbar^2}\Big)^{3/2}\int dE~(E+CE)^{3/2}(1+2EC)^{-1}\frac{\partial f_{FD}}{\partial E}.
\end{equation}

Comparing Eq. \ref{sigma} to the form of the Drude model conductivity

\begin{equation}
    \sigma=-\frac{\varepsilon_0\omega_p^2}{-i\omega+\gamma},
\end{equation}

\noindent we find that the plasma frequency is then, as given by Eq. \ref{PF} in the main text,

\begin{equation}
\label{PF1}
    \omega_p^2=\frac{e^2}{3m_0^*\varepsilon_0\pi^2}\Big(\frac{2m_0^*}{\hbar^2}\Big)^{3/2}\int dE~(E+CE)^{3/2}(1+2EC)^{-1}\frac{\partial f_{FD}}{\partial E}.
\end{equation}

Eq. \ref{PF1} is also derived in Ref. \citen{Guo2016Ultrafast} using a different method, but with the same conclusion. When $C = 0$, this expression becomes the traditional expression for the plasma frequency,

\begin{equation}
\label{Classic_PF}
    \omega_p^2=\frac{e^2n}{\varepsilon_0m_0^*}.
\end{equation}

For ITO the nonparabolicity factor was found to be $C=0.4191$ (eV \textsuperscript{-1}) \citep{Liu2014Quantification} and the effective mass was found to be  $m_0^*=0.4 m_e$ \cite{Wang2019Extended}, where $m_e$ is the free-electron mass.

\subsection*{Section 2: Validation for Neglecting the Athermal Electron Distribution}
\label{scattering_time}

In this section, we justify why the athermal electrons do not need to be accounted for explicitly in the two-temperature model for ITO (Eq. \ref{TTM}) that we introduced in the main text. Our theoretical analysis follows Ref. \citen{Fann1992Electron}, and we find that since the Fermi energy of ITO and the photon energy of our laser pulses are approximately the same,  $E_F\approx\hbar\omega_{pulse}\approx 1$ eV, the thermalization time is negligible compared to the pump pulse duration, and therefore we can neglect the athermal electrons in our model. This is not the case for materials with higher Fermi energies, like gold, to which the extended TTM is commonly applied (for example see Refs. \citen{Voisin2001Ultrafast,Carpene2006Ultrafast,Fann1992Electron}). 

\vspace{10pt}

Recall that the electron distribution function can be written as a sum of the thermal $f_{FD} (T_e,t)$ and athermal $\delta f(E,t)$ components as in Eq. \ref{delta}. The athermal electrons are the result of the ultrafast pulse perturbing the conduction electrons out of thermal equilibrium. As the pulse irradiates the sample, the electronic distribution deviates further from a thermal distribution. The electrons eventually become thermalized via collisions resulting in a Fermi-Dirac distribution at an increased temperature. The thermalization process is thus described by $\delta f(E,t)\rightarrow 0$ and $f_{FD} (T_{e,0},t=0)\rightarrow f_{FD} (T_{e,1}>T_{e,0},t)$. 

\vspace{10pt}

We begin by assuming that the lifetime of an excited electron (of energy $E$) in a degenerate system due to both elastic and inelastic electron-electron collisions is given by Fermi-liquid theory under the random-phase approximation as \cite{Fann1992Electron,Nozieres1999theory}

\begin{equation}
    \tau=\frac{128}{\pi^2\sqrt{3} \omega_p}\Big(\frac{E_F}{E-E_F}\Big)^2\approx\frac{7.5}{\omega_p}\Big(\frac{E_F}{E-E_F}\Big)^2,
\end{equation}

\noindent where $\omega_p$ is the plasma frequency. In the relaxation time approximation, we have

\begin{equation}
    \frac{\partial \delta f}{\partial t}\approx\frac{\delta f}{\tau}\approx \delta f \frac{\omega_p}{7.5}\frac{(E-E_F)^2}{E_F^2},
\end{equation}

\noindent and therefore

\begin{equation}
    \delta f(E,t)=\delta f(E,t=0)\exp\Big[-\frac{\omega_p}{7.5}\frac{(E-E_F)^2}{E_F^2}~t\Big],
\end{equation}

\noindent where $\delta f(E,t=0)$ is the initial athermal distribution. Here we assume that the perturbation  $\delta f(E,t=0)$ is instantaneous which, physically, is not the case. The perturbation will occur gradually as the pulse propagates through the sample. However, this assumption should not affect the conclusion of the following analysis, as we are only interested in how long it takes for conduction electrons to thermalize (on average).
\vspace{10pt}

\begin{figure}
\centering
\includegraphics[width = 3in]{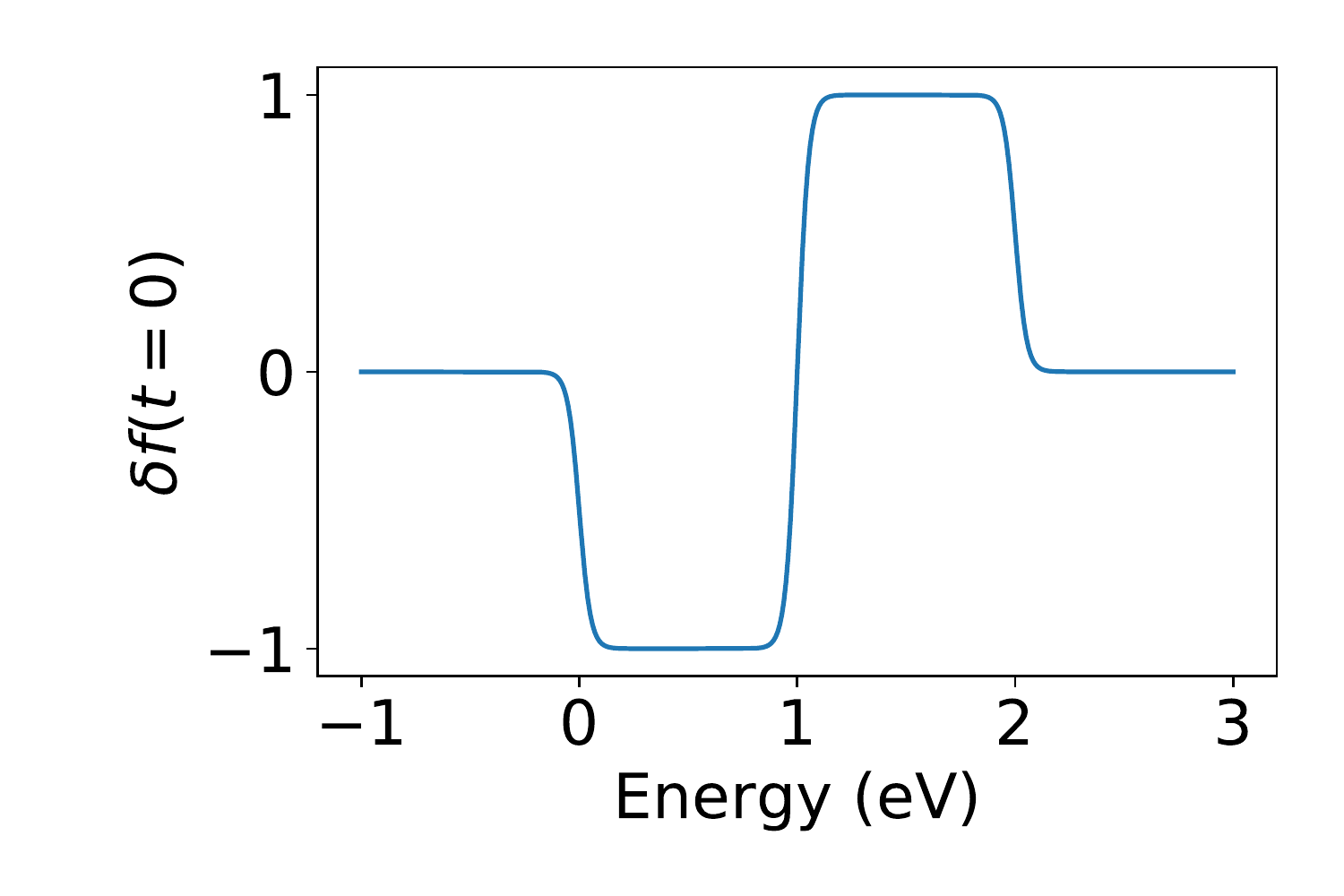}
\caption{Initial athermal electron distribution as a function of energy (above conduction band minimum) given by Eq. \ref{athermal}.}
\label{fig9}
\end{figure}

The initial distribution is described by the simplified model described in the main text (see discussion accompanying Fig 2), where electrons below the Fermi energy ($E_F - E_{pump}\rightarrow E_F$) are excited above the Fermi energy ($E_F\rightarrow E_F + E_{pump}$) and is thus given by \cite{Carpene2006Ultrafast}

\begin{equation}
\label{athermal}
    \delta f(E,t=0) = A(f_{FD}(E-E_{pump})[1-f_{FD}(E)]-f_{FD}(E)[1-f_{FD}(E+E_{pump})])  , 
\end{equation}

\noindent where $A$ is a constant amplitude which depends on the optical fluence of the pump pulse, and where we use $T_{e,0}=$ 300 K for the evaluation of $f_{FD}(E)$ which, as we will see, gives an upper bound for the athermal decay time. This function is plotted in Fig. \ref{fig9} for $E_F=E_{pump}=1.0$ eV and $A=1.0$. We also plot $f_{FD} + \delta f(t=0)$ in Fig. \ref{physics}b of the main text.

The energy density stored in the initial athermal distribution is given by

\begin{equation}
    U=\int_0^\infty D(E)\delta f(E,t=0)E~dE,
\end{equation}

\noindent where $D(E)$  is the density of states, which for non-parabolic bands can be derived using Eq. \ref{nparabolic}, and is given by \cite{Wang2019Extended}

 \begin{equation}
 \label{DOS}
     D(E)=\frac{1}{2\pi^2}\Big(\frac{2m_0^*}{\hbar^2}\Big)^{3/2}(E+CE^2)^{1/2}(1+2CE).
\end{equation}

Energy lost from the athermal electrons will be passed primarily to the thermal electron subsystem with (time-dependent) temperature $T_e$ and therefore we write Eq. \ref{TTM}a of the main text as 
\begin{equation}
\label{TTMsource}
    C_e\frac{\partial T_e}{\partial t}=-G(T_e-T_l)+\frac{\partial U}{\partial t},
\end{equation}

\noindent where the last term is the energy source of the thermalized electrons which will henceforth be denoted by

\begin{equation}
\label{Source}
    S(t)=\frac{\partial U}{\partial t}=\frac{\partial}{\partial t}\int_0^\infty D(E)\delta f(E,t=0)E~dE.
\end{equation}

\begin{figure}
\centering
\includegraphics[width = 3in]{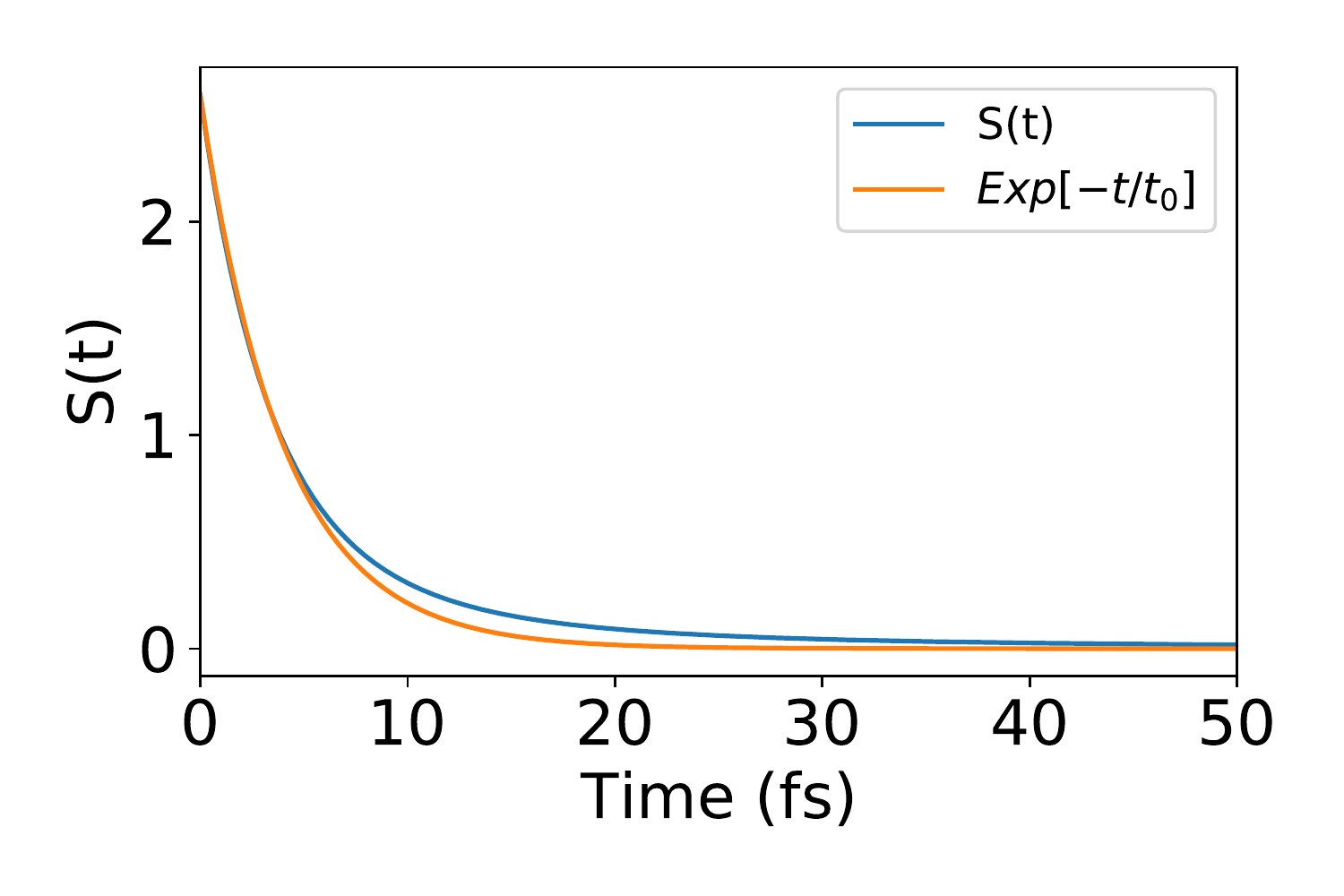}
\caption{Athermal electron energy density source function $S(t)$ for ITO. The blue line is the numerical integration of Eq. \ref{Source} and the yellow line is an exponential fit with $t_0=4.2$ fs.}
\label{fig10}
\end{figure}

We numerically integrate Eq. \ref{Source} (using Eq. \ref{athermal}), which we plot in Fig. \ref{fig10} where the blue line is the numerical integration, and the yellow line is an exponential fit with decay constant $t_0=$ 4.2 fs. This means that the athermal electrons decay with an effective decay time of approximately 4 fs and therefore the characteristic time of this process is negligible when compared to pulse duration (on the order of 100 fs).
\vspace{10pt}

As the thermalized electronic temperature $T_e$ increases, the thermalization time decreases even further. This is because there are more available high-energy states into which the athermal electrons can thermalize. In fact, for temperatures in the thousands of Kelvins, the thermalization time becomes sub-femtosecond.

\begin{figure}
\centering
\includegraphics[width = 3in]{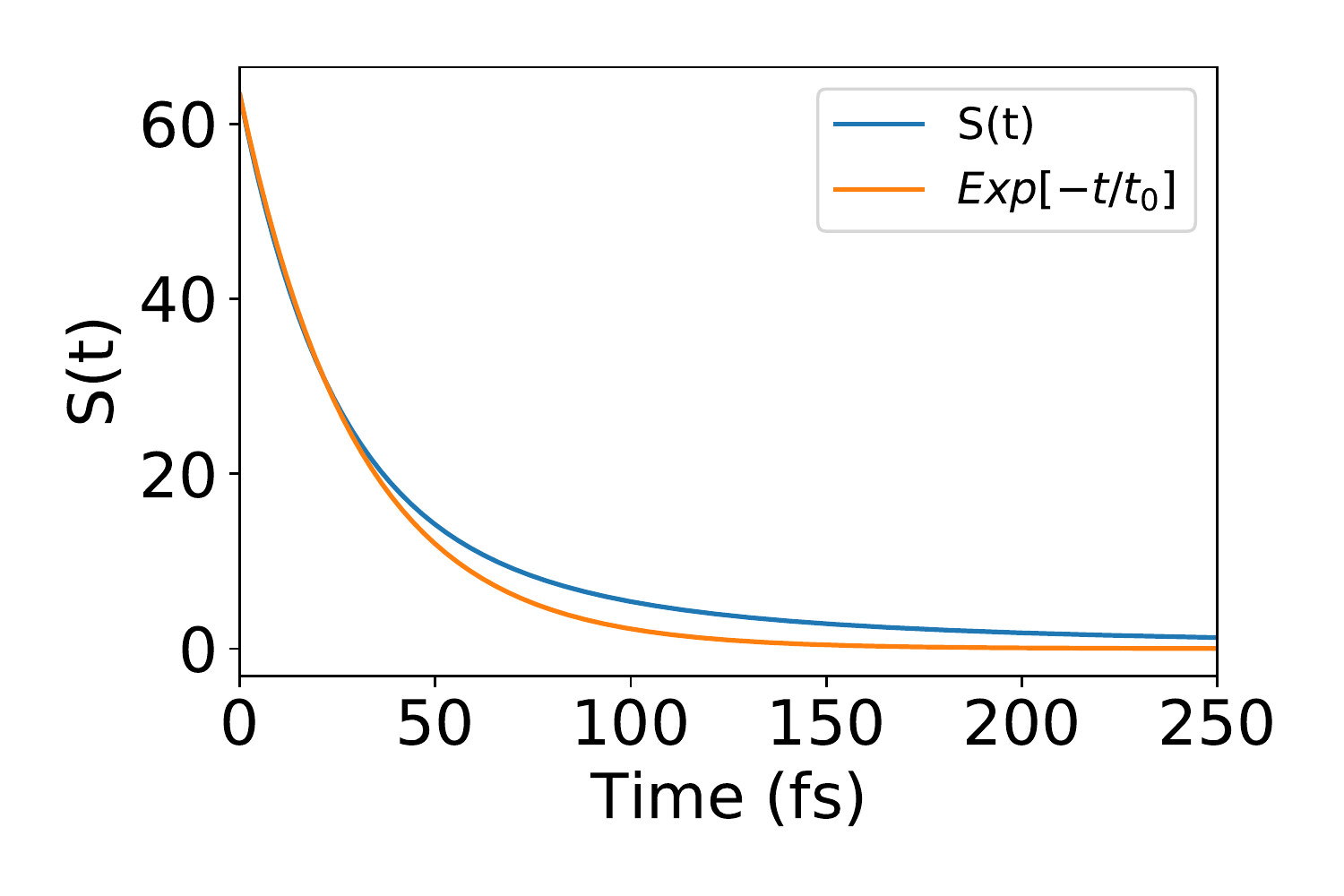}
\caption{Athermal electron energy density source function $S(t)$ for gold. The blue line is the numerical integration of Eq. \ref{Source} and the yellow line is an exponential fit with $t_0=30$ fs.}
\label{fig11}
\end{figure}

For reference, we repeat this calculation for gold with a room temperature Fermi energy of 5.53 eV and a plasma frequency of $\omega_p\approx 1.3\times10^{16}$  rad/s using the same pulse energy of 1 eV. $S(t)$, for this case, is plotted in Fig \ref{fig11} (blue) along with an exponential fit (yellow). Here we find the decay constant of the exponential fit to be $t_0$=30fs, which is comparable to typical ultrafast pulse widths of interest. Therefore, for gold (and indeed any other materials with $E_F>>E_{pump}$), the athermal electron distribution must be considered.

\subsection*{Section 3: Temperature Dependent Parameters of the Two Temperature Model}

In this section we give the formulas used for calculating the temperature-dependent parameters used in the two-temperature model Eqs. \ref{TTM}, including the heat capacities and the electron-phonon coupling coefficient. 

The electron heat capacity $C_e$ is calculated using \cite{Kittel2005Introduction} 

\begin{equation}
\label{EHC}
    C_e(T_e)=\int_0^\infty\frac{\partial f_{FD}(E,\mu,T_e)}{\partial T_e}D(E)E~dE,    
\end{equation}

\noindent where $D(E)$ is the density of states given in Eq. \ref{DOS}, and $\mu$ is the chemical potential (that will be discussed later). The lattice heat capacity is assumed constant at $C_l=2.54\times10^6$ Jm\textsuperscript{-3}K\textsuperscript{-1}  due to the small change in lattice temperature during the simulation \cite{Alam2018Large}. 

The electron-phonon coupling coefficient is calculated using \cite{Lin2008Electron-phonon,Alam2018Large}

\begin{equation}
\label{EPC}
    g_{ep}(T_e)=\frac{A_{ep}\pi\hbar k_B}{D(E_F)}\int_{-\infty}^\infty D^2(E) \frac{\partial f_{FD}(E,\mu,T_e)}{\partial T_e}dE,
\end{equation}

\noindent where $A_{ep}$ is the product of the electron-phonon mass enhancement parameter and the second moment of the phonon spectrum. As with Ref. \citen{Alam2018Large}, we take this to be a free parameter with $A_{ep}=5.25\times10^{-4}$ eV\textsuperscript{2}.

In the above formulas, we specify the dependence of the Fermi-Dirac distribution on the chemical potential $\mu$, which is also a temperature-dependent quantity. Numerically, its value can be extracted using the fact that the conduction band electron density is independent of temperature (in the absence of interband transitions), i.e., $n(T_e )=n(T_e=0)$ with

\begin{equation}
    n(T_e)=\int_0^{\infty}f_{FD}(E,T_e)D(E)~dE,
\end{equation}

and

\begin{equation}
    n(T_e=0)=\int_0^{E_F}D(E)~dE,
\end{equation}

\noindent where $E_F$ is the Fermi energy ($E_F\approx 1$ eV for ITO). Thus we extract $\mu(T_e)$ by minimizing the function $|n\big(\mu(T_e)\big)-n(T_e=0)|$ using a minimization algorithm.

Eqs. \ref{EHC}, \ref{EPC}, and \ref{PF1} can be integrated numerically. For our application, we fit these functions of $T_e$ to fifth-order, single-variable polynomials and use those in the TTM-FDTD code.

\renewcommand{\thefootnote}{\textit{\alph{footnote}}}

\subsection*{Section 4: Optimizing pump parameters for maximum index change}
\label{section:s5}

\begin{figure}
\centering
\includegraphics[width =\linewidth]{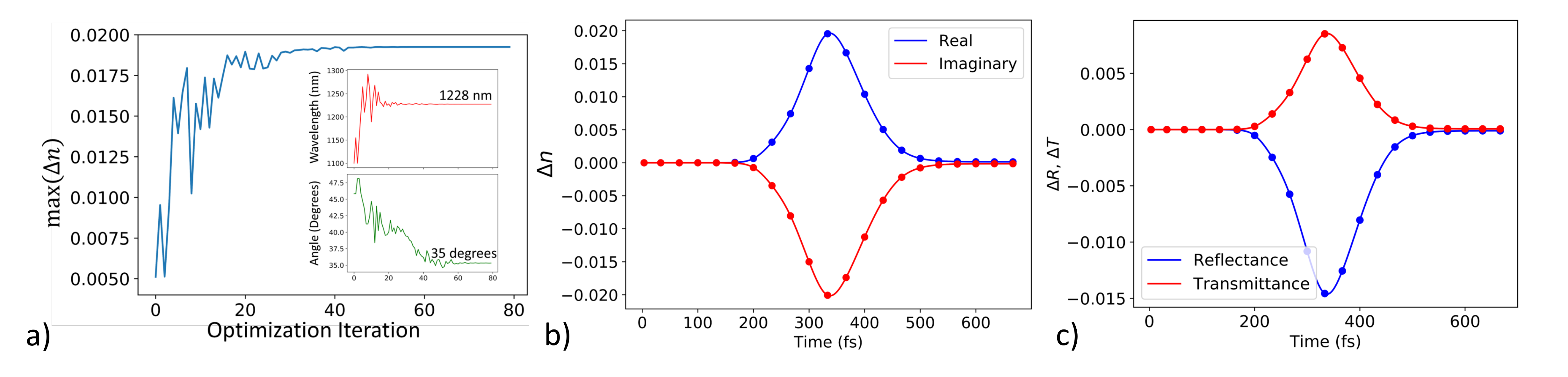}
\caption{a) Optimization of refractive index change by tuning pump pulse wavelength and incident angle using Nelder-Mead. The horizontal axis is the optimization iteration, and the vertical axis is the maximum change in the refractive index. The insets show how the central wavelength of the pump pulse and its angle of incidence (with respect to the interface normal) evolve during the optimization process. b) Complex refractive index change and c) reflectance and transmittance change as a function of time for the optimized pump pulse.}
\label{optimization}
\end{figure}

The main text was dedicated to validating and refining our hybrid TTM-FDTD approach. In doing so, we found that the properties of the incident excitation have a large effect on the response of ITO, a fact that is not necessarily surprising, but one that can be harnessed. In this section we use a Nelder-Mead optimizer to find the maximum change in refractive index, $\max[\Delta n(\textbf{r},t)]$\footnote{Note that $\Delta n$ presented in Fig. \ref{optimization}a and $n_2$ presented in Fig. \ref{fig1} of the main text are different values. The nonlinear refractive index $n_2$ is a space and time average of $\Delta n$ and is normalized by the intensity of the pump in the ITO layer, whereas $\Delta n(\textbf{r},t)$ is the instantaneous change of the refractive index.}, using three of the pump’s parameters: incident angle, central wavelength, and linear chirp parameter. Through previous trials, we find that by increasing the pulse width and laser intensity, we also increase the nonlinearity since this results in a larger energy density being deposited into the ITO. We therefore keep these values constant with a peak intensity of  $I_{pump}=$ 2 GW/cm\textsuperscript{2} and a full-width at half-maximum pulse width of $\tau_{pump}=$ 150 fs.

 \vspace{10pt}

 The optimization results are found in Fig. \ref{optimization}a, where the main plot shows the evolution of $\max[\Delta n]$ as the optimization progresses, and the insets show the progression of the wavelength (red line) and angle (green line). The optimization algorithm was run several times with different initial conditions so we are confident that we have found a global maximum. We find that the maximum change in the refractive index is on the order of 0.02 and occurs when $\lambda_0=$ 1228 nm, $\theta_{inc}=$ 35 degrees, and with negligible chirp (not shown). We found that the chirp of the pulse had very little effect on the nonlinearity. 
 
 \vspace{10pt}

\begin{figure}
\centering
\includegraphics[width =\linewidth]{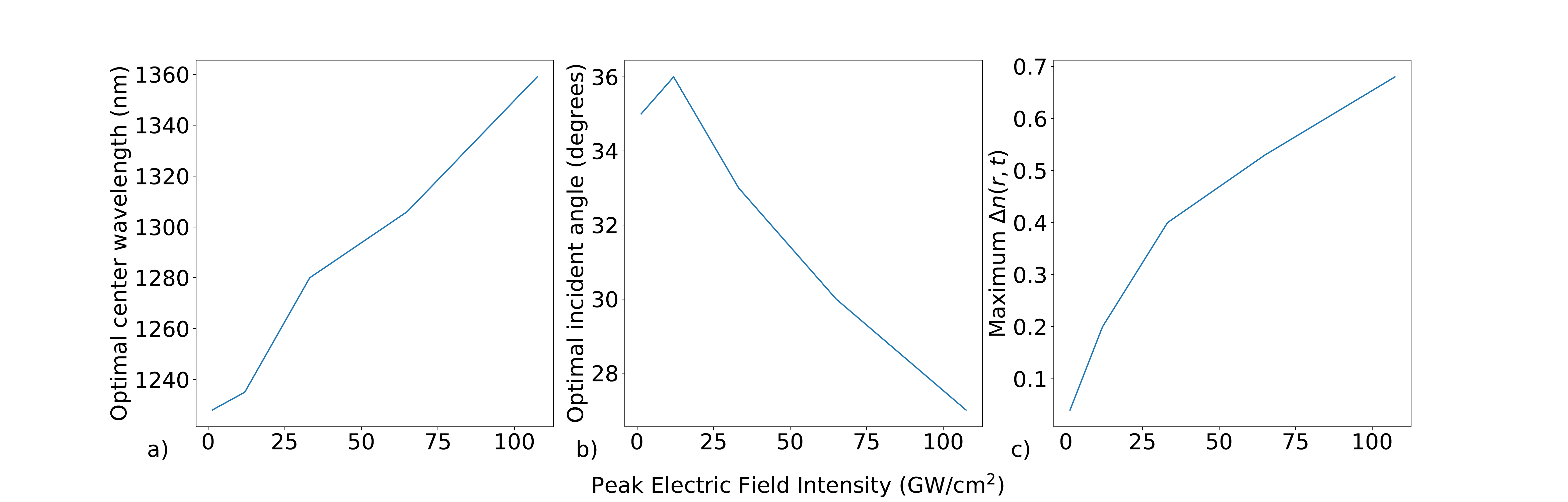}
\caption{The a) center wavelength, b) incident angle and c) resultant change in the refractive index of the optimized pump pulses found via Nelder Mead optimization, plotted here as a function of peak electric field intensity of the pulse. The pulses were modulated Gaussians with a pulse width of 150 fs.}
\label{optimization2}
\end{figure}



The results in Fig. \ref{optimization}b shows the time dependence of the optical response of the ITO film. We see that the real part of the refractive index increases and the imaginary part decreases by approximately 0.02. In Fig. \ref{optimization}c we plot the corresponding change in reflectance and transmittance, $\Delta R$ and $\Delta T$, respectively, as functions of time where we see that the transmittance increases during the pulse irradiation. This is not surprising as it is expected there would be less overall absorption in the ITO layer.

 \vspace{10pt}

We repeated this optimization for higher pulse intensities (up to 110 GW/cm\textsuperscript{2}). The results are summarized in Fig. \ref{optimization2}, where we show (a) the optimal central wavelength, (b) incident angle, and (c) the maximum of $\Delta n$ as functions of pump intensity. As expected, we see an increase in the maximum of $\Delta n$ with increasing intensity. We also see that the optimal incident angle remains relatively constant with a slight decrease with increasing intensity. Interestingly, we see an increase in the optimal wavelength with increasing intensity, which can be easily explained. As shown in Fig. \ref{pffig} of the main text, as the temperature is increased, the ENZ wavelength red-shifts. Therefore, in order for the pump pulse to remain in the ENZ spectral range, and thus to maximize the nonlinearity, we must tune our pulse to higher wavelengths for increased intensities.  

\subsection*{Section 5: Time-domain movies}
In Section 4\ref{section:s4} of the main text, we discuss the time-domain results of our model, and we introduce movies that can be found in the Supplementary Materials. We describe \textit{Movie1} and \textit{Movie1-T} in detail in Section 4\ref{section:s4}, and we will briefly describe the others here.

\vspace{10pt}

\textit{Movie2}, \textit{Movie3}, and \textit{Movie4} are also for a 1240 nm pulse, incident at 30 degrees, for intensities of 150, 50, and 10 GW/cm\textsuperscript{2}, respectively. In \textit{Movie2} and \textit{Movie3}, we see the aforementioned decreasing skin depth and change in transmission angle (in the ITO layer), however the reflectance remains high enough that the interference pattern is observed long after the pulse peak. In \textit{Movie4} there is no perceivable change in the optical response due to the low pump intensity.

\vspace{10pt}

\textit{Movie5} to \textit{Movie8} have the same intensity and incident angle as \textit{Movie1}, but with central wavelengths of 1160, 1200, 1280, and 1320 nm, respectively. We see the same effects as mentioned previously (angle shift and change in skin depth), however, we observe that the initial transmission angles are all different, which is because the initial permitivities are all different for the different wavelengths (as illustrated in Fig. \ref{pffig} of the main text). From these videos, we cannot observe a perceptible difference in the strength of the nonlinearity, however, we know from spectral results (see Fig. \ref{fig1} of the main text) that they do exist.

\vspace{10pt}

Finally, \textit{Movie9} and \textit{Movie10} have the same intensity and center wavelength as \textit{Movie1} but with incident angles of 0 and 60 degrees, respectively. In \textit{Movie9}, we see negligible change in the optical response as expected from Fig. \ref{fig1} of the main text (for 0 degrees). In \textit{Movie10}, the reflectance is strong throughout the simulation due to the large incident angle. However a perceptible shift in the transmitted angle in the ITO is visible.





\begin{acknowledgement}
    The authors would like to thank Compute Canada and
    Scinet for computational resources. J.B. and L.R. acknowledge the Natural Sciences and Engineering Research Council (NSERC) Vanier Canada Graduate Scholarships program and the Canada Research Chairs program for financial support. I.D.L., A.P.C. and L.C.H acknowledge financial support from CONACyT (Ciencia Básica) grant no. 286150. I.D.L. acknowledges the support of the Federico Baur Endowed Chair in Nanotechnology. A.C.L. acknowledges the Bundesministerium für Buldung und Furschung (German Federal Ministry of Education and Research) under the Tenure-Track Program, and the Deutsche Forschungsgemeinschaft (DFG, German Research
    Foundation) under Germany’s Excellence Strategy within the Cluster of Excellence PhoenixD (EXC 2122, Project ID390833453). The authors also thank the Max Planck Institute for the Science of Light for enabling this collaborative effort, through funding the travel of photonics researchers and their annual meeting.

    \end{acknowledgement}

\begin{suppinfo}

A supplemental information section is included with this paper which includes extra details supporting the main text. Movie files of our simulations are also included that show how the electric field amplitude and temperature fields evolve in time.

\end{suppinfo}

\bibliography{main.bib}

\end{document}